\documentclass[conference]{IEEEtran}
\IEEEoverridecommandlockouts
\usepackage{cite}
\usepackage{amsmath,amssymb,amsfonts}
\usepackage{algorithmic}
\usepackage[ruled, linesnumbered]{algorithm2e}
\usepackage{graphicx}
\usepackage{textcomp}
\usepackage{xcolor}
\usepackage{subcaption}
\usepackage{tabularx, booktabs}
\usepackage[braket,qm]{qcircuit}
\usepackage{multirow}
\usepackage{balance}

\def\BibTeX{{\rm B\kern-.05em{\sc i\kern-.025em b}\kern-.08em
    T\kern-.1667em\lower.7ex\hbox{E}\kern-.125emX}}
\begin{document}

\bstctlcite{IEEEexample:BSTcontrol}

\title{Quantum Interval Bound Propagation for Certified Training of Quantum Neural Networks}

\author{\IEEEauthorblockN{Emma Andrews, Nahyeon Kim, and Prabhat Mishra}
\IEEEauthorblockA{
\textit{University of Florida, 
Gainesville, FL, USA}}
}

\maketitle

\begin{abstract}
Quantum machine learning is a promising field for efficiently learning features of a dataset to perform a specified task, such as classification. Interval bound propagation (IBP) is a popular certified training method in classical machine learning, where the lower and upper bounds are tracked throughout the model. These bounds are used during training to ensure that the model is certified to predict the correct label even under adversarial perturbations. While IBP is successful in classical domain, there are limited certified training efforts in quantum domain. In this paper, we present quantum interval bound propagation (QIBP) to establish a certified training routine for quantum machine learning, certifying the accuracy of models under adversarial perturbations. We implement QIBP using both interval and affine arithmetic to explore the tradeoffs between the two implementations in terms of accuracy and other design considerations. Extensive evaluation demonstrates that the resulting certified trained models have robust decision boundaries, guaranteed to predict the correct class for the samples within the trained adversarial robustness bounds.
\end{abstract}


\section{Introduction}
\begin{figure}[t]
    \centering
    \includegraphics[width=\linewidth]{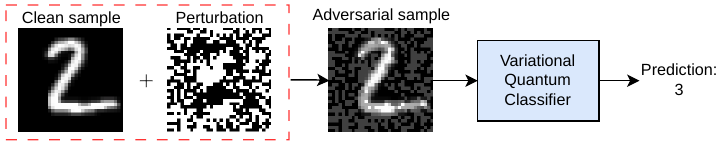}
    \caption{Example of an MNIST~\cite{lecun1998mnist} image of digit 2 being adversarially perturbed. While a human can still identify the digit 2, a quantum classifier will incorrectly predict it as 3.}
    \label{fig:introex}
    \vspace{-0.2in}
\end{figure}

Quantum machine learning (QML) has the potential to efficiently learn the features of a dataset compared to classical machine learning~\cite{schuld2015introduction}. This enables QML models to provide greater accuracy with fewer parameters compared to their classical counterparts. However, these models come with inherent risks, including being vulnerable to adversarial attacks~\cite{lu2020quantum}.
Attackers can carefully craft adversarial samples to intentionally get the model to change its resulting prediction from the correct class to an incorrect one. This is often achieved by adding a small amount of noise or an adversarial perturbation to a clean sample.  Figure~\ref{fig:introex} displays this phenomenon with the MNIST dataset~\cite{lecun1998mnist}, where a clean image of digit 2 is adversarially perturbed. To a human, the sample still appears as 2, whereas the classifier is incorrectly predicting it as 3.

There are several defense strategies against adversarial attacks for QML models. These defense frameworks aim to counteract the misclassifications from the adversarial samples through additional processing on either the input data samples or through additional model layers. Adversarial training is a popular training technique where the model itself or the defense framework are trained on adversarial samples from adversarial attacks. 
While these defense frameworks and training routines work well in combating adversarial attacks, they are not able to solve the core problem with the models causing these adversarial attacks to occur in the first place. When the model is trained, the decision boundaries are established on clean samples, and with adversarial training, these bounds are updated to become robust to the exposed adversarial samples. Adversarial training can lead to overfitting towards the adversarial attack patterns it was trained on and may not be effective against all possible adversarial samples. 

There is a critical need for developing training methods that can establish adversarially robust decision boundaries to ensure that adversarial samples are classified correctly without relying on adversarial training. Certified training is popular in   classical machine learning, where the decision boundaries are certified to always classify an adversarially perturbed sample correctly under a specified perturbation. Specifically, interval bound propagation (IBP) is a classical certified training technique that ensures input data samples remain within safe regions as they propagate through the layers in the model~\cite{gowal2019scalable, mao2024understanding}. During model training, data samples are turned into intervals with respect to a defined perturbation $\epsilon$. These intervals are computed throughout the model, tracking how the interval changes at each major point, such as after a layer. 
Training the model to improve the boundaries with respect to this certification criteria is carried out to ensure that the model creates and maintains robust decision boundaries.  

In this paper, we propose quantum interval bound propagation (QIBP) for certifying the training of quantum neural networks, primarily the variational quantum classifier (VQC) family of models for image datasets. \textit{To the best of our knowledge, this is the first attempt to establish a certified training method via IBP for QML models.} We develop a training routine for VQC that propagates the intervals using either interval arithmetic or affine arithmetic across stages, from feature mapping to measurement. Each layer processes the interval or affine term, resulting in final bounds that can be used in training. With this certified training routine, we are able to certify with tight bounds to the original test accuracy and explore tradeoffs resulting from training under interval arithmetic versus affine arithmetic.

Specifically, this paper makes the following contributions:
\begin{itemize}
    \item We propose a certified training technique using interval bound propagation for VQC models.
    \item We explore propagation of intervals through the VQC model using both interval and affine arithmetic.
    \item Results demonstrate the ability to tightly certify the VQC model's original test accuracy for a given perturbation $\epsilon$.
\end{itemize}

The remainder of the paper is organized as follows. Section~\ref{sec:bg} provides necessary background and surveys related efforts. Sections~\ref{sec:qibp-ia} and~\ref{sec:qibp-affine} describe QIBP using interval arithmetic and affine arithmetic, respectively. Section~\ref{sec:results} presents experimental results. Finally, Section~\ref{sec:conc} concludes the paper.
\section{Background and Related Work} \label{sec:bg}
In this section, we first provide necessary background on interval and affine arithmetic, QML, adversarial training, and certified training. Next, we survey related efforts  and their limitations to highlight our contributions.

\subsection{Interval and Affine Arithmetic}
Interval arithmetic is a mathematical technique for performing operations on intervals. An interval is defined as having an upper and a lower bound of the form $c=[a,b]$. We denote $\underline{c}$ to represent the lower bound ($a$) and $\overline{c}$ to represent the upper bound ($b$). A mathematical operation is performed on the interval itself, with linear, monotonic, or constant functions applying directly to the bounds of the interval to obtain the resulting interval. Nonlinear functions require searching the interval for the resulting minimum and maximum to establish the new interval. Additionally, a mathematical operation can be carried out with the intervals as the operands. For example, interval multiplication results in
\begin{equation}
    [a,b]\times[c,d]=\left[\min(ac,ad,bc,bd), \max(ac,ad,bc,bd)\right].
\end{equation}

The interval multiplication routine highlights the dependency problem in interval arithmetic. The dependency problem occurs when the intervals are not used in an independent form during computation, such as in multiplication. This leads to resulting intervals that are not tight to the model, potentially causing issues within the application use. To avoid these problems, affine arithmetic can be used, where the intervals are instead expressed as a linear combination of the terms with noise coefficients to handle errors. While this does provide tighter bounds compared to interval arithmetic, it often comes at the cost of additional, expensive computations.

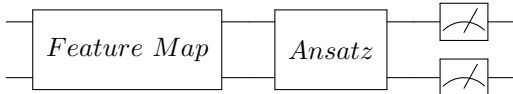
\begin{figure}[h]
    \vspace{-0.1in}
    \centering
    \centerline{\Qcircuit @C=1em @R=.7em {
        & \multigate{1}{Feature\ Map} & \qw & \multigate{1}{Ansatz} & \qw & \meter & \qw\\
        & \ghost{Feature\ Map} & \qw & \ghost{Ansatz} & \qw & \meter & \qw
    }}
    \vspace{-0.05in}
    \caption{Common structure of QML models.}
    \label{fig:qnn}
        \vspace{-0.1in}
\end{figure}

\subsection{Quantum Machine Learning}
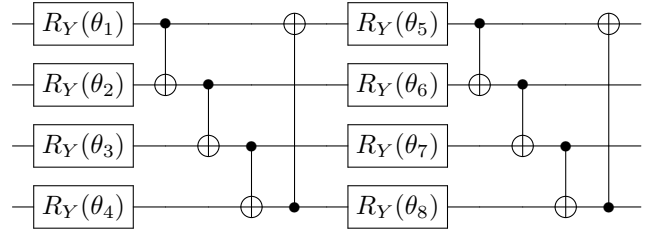
\begin{figure}[h]
    \vspace{-0.1in}
    \centering
    \centerline{\Qcircuit @C=0.8em @R=.7em {
        & \gate{R_Y(\theta_1)} & \ctrl{1} & \qw & \qw & \targ & \qw & \gate{R_Y(\theta_5)} & \ctrl{1} & \qw & \qw & \targ & \qw\\
        & \gate{R_Y(\theta_2)} & \targ & \ctrl{1} & \qw & \qw & \qw & \gate{R_Y(\theta_6)} & \targ & \ctrl{1} & \qw & \qw & \qw\\
        & \gate{R_Y(\theta_3)} & \qw & \targ & \ctrl{1} & \qw & \qw & \gate{R_Y(\theta_7)} & \qw & \targ & \ctrl{1} & \qw & \qw\\
        & \gate{R_Y(\theta_4)} & \qw & \qw & \targ & \ctrl{-3} & \qw & \gate{R_Y(\theta_8)} & \qw & \qw & \targ & \ctrl{-3} & \qw\\
    }}
    \caption{Two layers of a QML ansatz. This ansatz features a layer consisting of $R_Y$ gates parameterized by $\theta$, followed by entanglement through pairs of CNOT gates. This layer is repeated for the second layer of the ansatz.}
    \label{fig:ansatz}
        \vspace{-0.1in}
\end{figure}

Figure~\ref{fig:qnn} provides an overview of the basic structures typically found in QML models. These are the feature map, the ansatz, and measurement. The feature map is responsible for embedding the data into the quantum circuit of the QML model. The ansatz carries out the QML model computations, consisting of parameterized and entangling gates. The parameters of the parameterized gates are updated over the course of training and act as the weights of the model, like in classical machine learning. An example two-layer ansatz using parameterized $R_Y$ gates is shown in Figure~\ref{fig:ansatz}.

\subsection{Adversarial Training}
Adversarial training is the process of training the VQC model using both clean and adversarial samples~\cite{lu2020quantum}. In these training methods, the loss function of the model can be updated from the traditional loss function with the goal to maximize the chances that adversarial samples are identified through their clean features rather than their adversarial features.
Adversarial training requires creation of adversarial samples during training. Often, this is done using state-of-the-art adversarial attacks, such as fast gradient sign method (FGSM)~\cite{goodfellow2015explaining} and projected gradient descent (PGD)~\cite{madry2019deep}, which utilize the gradients from the model to craft adversarial samples such that the model will purposefully predict the wrong class label. However, training on these crafted adversarial samples can potentially create an overfitting to these specific attacks, only protecting against the adversarial attacks the model encountered during training and reduced effectiveness against unknown adversarial attacks.

\subsection{Certified Training}
Certified training is the technique of training a model with the goal of making the decision boundaries the model learns adaptive to perturbations. This means that a model's decision boundaries are robust up to $\epsilon$ such that if an original data sample were perturbed by $\epsilon$ by an adversarial attack, the resulting adversarial sample will still be contained within the correct class's decision boundary. Thus, the model will be able to predict the adversarial sample as the correct class.

Lipschitz regularization is a training technique where the loss function includes Lipschitz bounds to regularize the term~\cite{pauli2022training}. This regularization can be expressed as 
\vspace{-0.05in}
\begin{equation} \label{eq:lip}
    \Vert f(x)-f(y)\Vert\leq L\Vert x-y\Vert,
    \vspace{-0.05in}
\end{equation}
where $L$ is taken as a small value to prevent adversarial samples from flipping the class label from the correct label to an incorrect one. However, this may require adversarial training to find an efficient $L$ for a class of adversarial attacks.

\begin{figure}
    \centering
    \includegraphics[width=\linewidth]{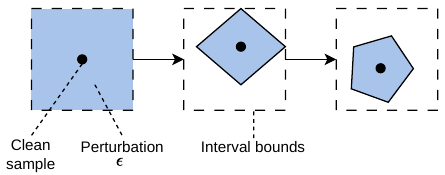}
    \caption{An  example of interval bound   through model layers.}
    \label{fig:intbounds}
        \vspace{-0.1in}
\end{figure}

In classical machine learning, IBP establishes intervals on the original data samples perturbed by some $\epsilon$ value~\cite{gowal2019scalable}. With the established interval $[x-\epsilon,x+\epsilon]$, the interval is propagated through the layers of the models, performing interval arithmetic on the intervals to establish the new interval after each layer, as shown in Figure~\ref{fig:intbounds}. This arithmetic differs depending on the linearity and monotonicity of the model layer, with linear and monotonic layers lending to fast operations on the intervals. Once the interval has been propagated throughout the entire model, the resulting intervals are taken as the worst case upper and lower bounds per class logit the model is classifying. The model is then trained to optimize these bounds such that the lower bound of the correct class is higher than the upper bounds of all the incorrect classes. This effectively establishes that in the worst case under $\epsilon$ perturbation, the model will always be able to classify the sample with the correct class as it is not possible for the incorrect class to have a logit value greater than the correct class.

\subsection{Related Work}





Adversarial training involves training the model with the adversarial samples and may include some additional defense mechanisms. This includes generative defenses, where classical autoencoders can be adversarially trained to reconstruct and purify the adversarial sample of its adversarial perturbations~\cite{khatun2025classical}. Non-defense methods train the model on the adversarial samples directly, in addition to the clean ones~\cite{wendlinger2024comparative}. While these defenses and adversarial training of the model can improve the accuracy of the model under adversarial samples, these defenses are often only available at inference time once the model has already been trained. Additionally, most make no guarantee about the input landscape that they are protecting against, making it possible for attacks to still work even with the defenses in place. We propose certified training for QML models that provides the guarantee as the decision boundaries are trained to account for perturbations up to $\epsilon$.

Lipschitz regularization is a training technique that can improve the robustness of model bounds~\cite{berberich2024training, wendlinger2024comparative}. While Lipschitz regularization can certify models and provide robustness against adversarial samples, determining $L$ in Equation~\ref{eq:lip} that is most appropriate for the model and dataset is difficult. In addition, Lipschitz regularization for QML models has only been devised for re-uploading models~\cite{perez-salinas2020data}.

There are also recent efforts in formal verification that can provide the robustness guarantees for an already trained model, determining the $\epsilon$ to which a model is robust. Prior work like VeriQR~\cite{lin2024veriqr} provide a verification on model robustness, providing exact verification of a perturbation locally or globally within the model. Other formal verification techniques involve semantic abstraction of the model through interval arithmetic to verify interval bounds on trained models~\cite{assolini2025formal}. While this tracks an interval through the model similar to IBP, formal verification can only verify already trained models under abstract semantics on the intervals. In contrast, our proposed QIBP can train the models to account for both clean and perturbed samples. This ensures that the model learns robust decision bounds for both clean and adversarial samples.

While certified training has been explored in classical machine learning, to the best of our knowledge, this paper is the first attempt at utilizing IBP and exploring tradeoff between interval arithmetic and affine arithmetic for QML models. Our certified  training methodology provides the ability  to learn robust decision boundaries without overfitting to specific adversarial attacks using interval arithmetic (Section~\ref{sec:qibp-ia}) as well as affine arithmetic (Section~\ref{sec:qibp-affine}).
\section{QIBP with Interval Arithmetic} \label{sec:qibp-ia}
\begin{figure*}
    \centering
    \includegraphics[width=\linewidth]{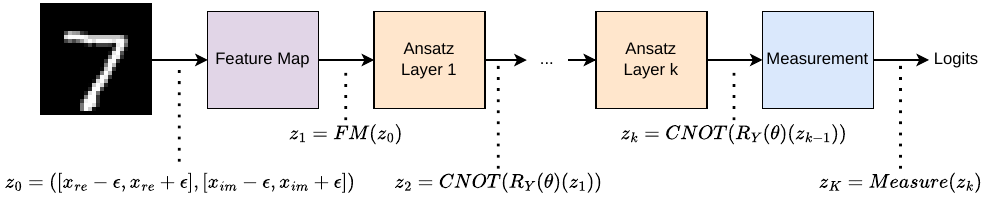}
    \caption{Overview of QIBP. In each layer, the interval shifts due to the layer operations, adjusting the worst case bounds. Once the interval is propagated through all layers, it results in the logits $z_K$, which are used to predict the class label. QIBP optimizes the interval logits so that the lower bound of the correct class logit is greater than the upper bounds of all incorrect class logits.}
    \label{fig:arch}
    \vspace{-0.15in}
\end{figure*}

Figure~\ref{fig:arch} provides an overview of our quantum interval bound propagation (QIBP) framework using interval arithmetic for QML models. In this section, we first formulate the problem. Next, we discuss how to incorporate intervals at different stages of the QML models. Finally, we describe training loss functions for robust decision boundaries.

\subsection{Problem Formulation}
We first establish some preliminary notations. The input data samples must be converted into a starting interval under a given perturbation $\epsilon$ that will be propagated through the $n$ qubit QML model. The input data sample with $N=2^n$ features is given by the vector $x=[x_0,x_1,\dots,x_N]$ and is perturbed by $\epsilon$ to establish the upper and lower bounds for the interval, resulting in $[x-\epsilon,x+\epsilon]$. We aim to optimize the model such that the decision boundaries are robust against $\epsilon$, meaning that a perturbation up to $\epsilon$ will not change the model's predicted class. 

The outputs of the layers, including from the feature map and measurement, are denoted with the general function
\vspace{-0.05in}
\begin{equation}
    z_k=U_k(z_{k-1}),
    \vspace{-0.05in}
\end{equation}
where $U_k$ is the unitary operation representing layer $k$ for $k=1,\dots,K$ layers, and $z_k$ the output from layer $k$. Therefore, for all layers except measurement, $z_k$ will represent the output statevector of the layer. After measurement, $z_K$ will represent the output logits, as outlined in Section~\ref{sec:ia-measure}.

As quantum computing handles complex numbers, we establish the intervals to have a real and an imaginary interval, making up a single interval term. When operations are performed on the interval, the interval arithmetic follows the logic in complex arithmetic. For example, complex multiplication between two complex number $a+bi$ and $c+di$ results in the final value of
\vspace{-0.05in}
\begin{equation}
    (a+bi)(c+di)=ac-bd+(ad+bc)i.
    \vspace{-0.05in}
\end{equation}
When used for intervals, multiplication will have to carry over both the real and imaginary intervals appropriately.

\subsection{Embedding Input Interval into Feature Map}
With the input data sample established into the complex intervals $z_0=([x_{re}-\epsilon,x_{re}+\epsilon], [x_{im}-\epsilon,x_{im}+\epsilon])$, this is propagated through the feature map. For image datasets and other high-dimensional data, amplitude embedding is used to embed the input data into the quantum circuit representing the model. Amplitude embedding prepares the state of the quantum circuit such that $N$ features are prepared into the statevector of $n$ qubits. This is also expressed as
\vspace{-0.05in}
\begin{equation}
    |\psi\rangle=\sum^{N}_{i=0}x_i|i\rangle,
    \vspace{-0.05in}
\end{equation}
where $x$ is the given input data sample and $|\psi\rangle$ the resulting prepared state~\cite{schuld2018supervised}. As this is a direct state preparation with no additional computation on $x$ directly, the interval is denoted as the amplitudes of the corresponding basis states as 
\vspace{-0.05in}
\begin{equation}
    z_1=I_{0,1}|0\dots00\rangle+I_{1,1}|0\dots10\rangle+\dots+I_{N,1}|1\dots11\rangle,
    \vspace{-0.05in}
\end{equation}
where $I_{j,1}=([x_{j,re}-\epsilon,x_{j,re}+\epsilon], [x_{j,im}-\epsilon,x_{j,im}+\epsilon])$ for the $j$th feature and $k=1$.

For additional feature maps, the interval arithmetic can be extended to support propagation through the feature map, such as with angle embedding. In these cases, care must be taken when propagating the interval, as the monotonicity of the unitary matrices must be handled properly with the intervals.

\subsection{Propagation through Ansatz Layers}
Following the feature map are layers of the ansatz, which are typically populated with single-qubit parameterized rotation gates and entangling gates. We consider first a layer of single-qubit parameterized rotation gates affecting the qubits, such as the first four $R_Y(\theta_i)$ gates for $i=1,\dots,4$ in Figure~\ref{fig:ansatz}. The angles $\theta_i$ for these gates are given by the current weights of the model. Therefore, the matrices of these gates will be evaluated to a constant value consisting of real or complex numbers, depending on the gate. 

As this now takes the shape of the calculation $y=Wx$, we can use the shortform linear layer calculations presented in classical IBP~\cite{gowal2019scalable} to derive the propagated intervals. Therefore, for input interval $I_{j,k-1}$ to layer $k$ results in the output interval $I_{j,k}$, the lower and upper bounds are derived as
\vspace{-0.05in}
\begin{gather}
    \underline{I}_{j,k}=M^+\underline{I}_{j,k-1}+M^-\overline{I}_{j,k-1}\\
    \overline{I}_{j,k}=M^+\overline{I}_{j,k-1}+M^-\underline{I}_{j,k-1},
    \vspace{-0.05in}
\end{gather}
where $\underline{I}_{j,k}$ is the lower bound, $\overline{I}_{j,k}$ is the upper bound, $M^+=\max(W,0)$, and $M^-=\min(W,0)$. Note that with complex interval multiplication, it must obey the multiplication routines for regular complex numbers.

The entangling gates also perform specific operations between two qubits, and thus in interval form, between amplitude intervals corresponding to the qubits. CNOT gates, typically used as the entangling gates in VQC models for image datasets, will permute the intervals to different amplitudes. This permutation does not affect the values inside of the tracked intervals, for both the real and imaginary parts. Thus, the CNOT operation will swap the intervals around in the current interval amplitude vector for the basis states accordingly.

\subsection{Measurement for Resulting Logits} \label{sec:ia-measure}
Measuring the qubits after the ansatz finishes its operations produces the output of the model. For classification, this is typically done with the expectation value using the Pauli-Z operation, producing logits representing the classification outcome of which label to predict the input data sample as. The expectation value first uses the probability measurement to obtain the greatest square amplitude of the statevector $z_k$ after $k$ layers. This results in
\vspace{-0.05in}
\begin{equation} \label{eq:ia-prob}
    z_{k+1}=z_{k,re}^2+z_{k,im}^2,
    \vspace{-0.05in}
\end{equation}
which produces a new real-valued interval $z_{k+1}$.

With the probability measurement, the expectation value requires additional processing to produce the logits. For qubit~$i$, $\langle Z_i\rangle=P(q_i=0)-P(q_i=1)$. Because the probabilities must sum to 1,  $P(q_i=0)+P(q_i=1)=1$, we can rewrite as 
\vspace{-0.05in}
\begin{equation} \label{eq:expval}
    z_K=2z_{k+1,P(q_i=0)}-1.
    \vspace{-0.05in}
\end{equation}
This gives the final intervals for use in the loss functions.

\subsection{Training Loss Functions for Robust Decision Boundaries} \label{sec:loss}
For both interval and affine arithmetic, the loss function used during training is the same. Once the final interval bounds for the logits on each possible class are produced, they are used in a two-term loss function. The first term of the loss function is the traditional classification objective to maximize predicting the labels as the correct label. This is done using cross entropy with $\mathcal{L}_{CE}(z,y_{true})$, where $z$ is the logits measured from the model and $y_{true}$ the true labels for each sample.

The second term in the loss function is the main objective for optimizing for robust decision boundaries. From the bounded logits, the worst-case logits is selected, expressed as
\vspace{-0.05in}
\begin{equation} \label{eq:logits}
    \hat{z}_K=\begin{cases}
        \overline{z}_{K,y} & \text{if}\ y \neq y_{\text{true}}\\
        \underline{z}_{K,y} & \text{otherwise}
        \vspace{-0.05in}
    \end{cases},
\end{equation}
where $\hat{z}_K$ is the worst case logits resulting after propagating through the entire model with the bounds chosen based on if the logit corresponds to the correct class or not. The worst case logits are used in cross entropy against the class label.

Thus, the resulting loss function with both terms is
\vspace{-0.05in}
\begin{equation} \label{eq:loss}
    \mathcal{L}=\kappa(\mathcal{L}_{CE}(z,y_{true}))+(1-\kappa)(\mathcal{L}_{CE}(\hat{z}_K,y_{true}))
    \vspace{-0.05in}
\end{equation}
where $\hat{z}_K$ are the bounded logits and $\kappa$ is a hyperparameter scheduled to control the impact of each term on the overall loss during training.
An alternative is margin loss~\cite{liu2016large}. The worst-case margin is defined as
\vspace{-0.05in}
\begin{equation}
    \text{margin} = l_y - \max_{j \neq y} u_j,
    \vspace{-0.05in}
\end{equation}
where \(l_y\) is the lower bound of the true class and \(u_j\) are the upper bounds of the other classes. To enforce a minimum separation under worst-case perturbations, a fixed margin target (e.g., \(\gamma = 0.2\)) is used. The margin target \(\gamma = 0.2\) is chosen based on the empirical margin observed when training without the margin loss. Specifically, it corresponds to approximately 80\% of the typical margin value in that setting, which was found experimentally to provide stable and effective training. The corresponding hinge-style loss~\cite{zhang2021boosting} is
\vspace{-0.05in}
\begin{equation} \label{eq:marginloss}
    \mathcal{L}_{\text{robust}} = \max(0, \gamma - \text{margin}).
    \vspace{-0.05in}
\end{equation}
If the margin is greater than or equal to \(\gamma\), the loss is zero, meaning no penalty is applied. If the margin is less than \(\gamma\), the model is penalized proportionally to the gap \(\gamma - \text{margin}\). This formulation directly encourages larger separation between class bounds, which in turn improves certified robustness.

\section{QIBP with Affine Arithmetic} \label{sec:qibp-affine}
Affine arithmetic provides tighter  intervals compared to interval arithmetic, where the intervals are expressed as affine terms such as first order Taylor series. Once the intervals are past feature mapping, the ansatz layers are linear transformations, enabling efficient use of affine terms to handle the intervals. The affine term is expressed as
\vspace{-0.05in}
\begin{equation}
    \hat{x}=x_0+x_1\varepsilon_1+\dots+x_n\varepsilon_n,
    \vspace{-0.05in}
\end{equation}
where $\varepsilon_i$ are the noise coefficients for each term in $x$ and are on the range $[-1,1]$~\cite{comba1993affine, defigueiredo2004affine}. While calculations on the affine term take more time compared to interval arithmetic, the construction of affine terms allows for a polynomial approximation to be created instead of a rectangle, tightening the upper and lower bounds of the resulting interval. 

\subsection{Embedding Input Affine Term into Feature Map}
The feature map is the first step of establishing the interval $[x-\epsilon,x+\epsilon]$ into the quantum circuit as an affine term. The exact formation of the affine term as a result of the feature map depends on the feature map used, like in interval arithmetic. The amplitude embedding prepares the input data $x$ into the statevector $|\psi\rangle$. To prepare the interval $[x-\epsilon,x+\epsilon]$ into the affine term, $x$ is used as the center terms in $\hat{x}$, with the noise coefficients $\varepsilon_i$ being set to the given $\epsilon$ for the real component. The imaginary affine term is similarly set if $x$ is complex, otherwise it is set to zero. Thus, this can be expressed as
\vspace{-0.05in}
\begin{equation}
    \hat{z}_1=(x_{0,re}+\sum_{i=1}^{n}x_{i,re}\epsilon,x_{0,im}+\sum_{i=1}^{n}x_{i,im}\epsilon).
    \vspace{-0.05in}
\end{equation}

\subsection{Propagation through Ansatz Layers}
The ansatz layers act as linear transformations with the parameterized rotation gates and swaps for the CNOTs. The resulting affine term $\hat{z}_{1}$ from the feature map is the input affine term to the first layer. For the parameterized rotation gates, these act as a linear transformation on the affine terms, which is equivalent to scalar affine multiplication.

In scalar affine multiplication, the constant scalar is multiplied throughout the affine term, such that
\vspace{-0.05in}
\begin{equation}
    \alpha\hat{x}=\alpha\left(x_0+\sum_{i=1}^nx_i\varepsilon_i\right).
    \vspace{-0.05in}
\end{equation}
For real valued matrices, such as parameterized $R_Y$ gates, scalar affine multiplication is carried out between the resulting real value matrix representing $R_Y(\theta)$ on the real and imaginary affine terms.

In the cases where the multiplication involves real and imaginary components, such as a parameterized $R_X$ gate, complex multiplication is required for addressing both the real and imaginary affine terms. Thus, multiplication between two affine terms ($\hat{a}$ and $\hat{b}$) is necessary, and can be expressed as 
\vspace{-0.05in}
\begin{gather}
    \hat{a}\times\hat{b}=\left(a_0+\sum_{i=1}^na_i\varepsilon_i\right)\times\left(b_0+\sum_{i=1}^nb_i\varepsilon_i\right) \nonumber \\
    =a_0b_0+\sum_{i=1}^n(a_0b_i+b_0a_i)\varepsilon_i+\sum_{i=1}^na_i\varepsilon_i+\sum_{i=1}^nb_i\varepsilon_i.
    \vspace{-0.05in}
\end{gather}
It can be extended to handle multiplication between complex numbers, as divided into their real and imaginary affine terms.

For each layer of parameterized gates, either scalar or affine multiplication is carried out depending on the values in the unitary matrix of the parameterized gate. This results in new real and imaginary affine terms, representing the propagation through the layer's parameterized gate. After the parameterized gates, there are entangling gates. Like with interval arithmetic, if these entangling gates are CNOTs, the corresponding affine terms in the statevector are swapped. The layer operations, depending on the structure of the model's layers, results in 
\vspace{-0.05in}
\begin{equation}
    \hat{z}_k=U_k\hat{z}_{k-1}
    \vspace{-0.05in}
\end{equation}
after layer $k$ within the ansatz.

\subsection{Measurement for Resulting Logits}
Measurement first begins with taking the probabilities of the amplitudes of the final statevector. This is done by squaring the resulting affine term from the final ansatz layer. The final probability value after $k$ layers is defined as
\vspace{-0.05in}
\begin{equation} \label{eq:prob-aff}
    \hat{z}_{k+1}=\hat{z}_{k,re}^2+\hat{z}_{k,im}^2,
    \vspace{-0.05in}
\end{equation}
similar to the probability measurement (Equation~\ref{eq:ia-prob}) in interval arithmetic. However, squaring an affine term results in a nonlinear transformation. With nonlinear functions, the affine term must be approximated instead of computed directly as with linear transformations. A nonlinear transformation thus results in an additional noise term being added to the affine term to represent this approximation, expressed as
\vspace{-0.05in}
\begin{equation}
    \hat{x}=x_0+\sum_{i=1}^nx_i\varepsilon_i+x_m\varepsilon_m,
    \vspace{-0.05in}
\end{equation}
where $x_m\varepsilon_m$ is the additional noise term. As this can result in many additional terms and an over-approximation, we utilize the Chebyshev approximation, where a residual term $r$ is used instead to bound all additional noise terms. Thus, we calculate the squaring and produce an affine term of the form
\vspace{-0.05in}
\begin{equation}
    \hat{x}=x_0+\sum_{i=1}^nx_i\varepsilon_i+r.
    \vspace{-0.05in}
\end{equation}


For expectation value measurements instead of probabilistic measurements, the affine term must undergo additional propagation. We start the expectation value propagation by propagating through the probability measurement as described in Equation~\ref{eq:prob-aff}, resulting in the single affine term $\hat{z}_{k+1}$. We utilize Equation~\ref{eq:expval}, where the probability of the zero basis terms is taken for each qubit index. This operation is performed on the affine term directly, resulting in
\vspace{-0.05in}
\begin{equation}
    \hat{z}_K=2\hat{z}_{k+1,P(q_i=0)}-1.
    \vspace{-0.05in}
\end{equation}
With this affine term, it is converted into an interval with an upper and lower bound to represent the logits. The bounds are established by adding and subtracting $x_0$ in $\hat{z}_K$ by the noise coefficients plus the residual, or in notation,
\vspace{-0.05in}
\begin{equation}
    z_K=\left[ x_0-\sum_{i=1}^n|x_i\varepsilon_i|-r, x_0+\sum_{i=1}^n|x_i\varepsilon_i|+r\right].
    \vspace{-0.05in}
\end{equation}
With the logits in interval form, the resulting $z_K$ can be used in the loss functions as defined in Section~\ref{sec:loss}.

\section{Experimental Results} \label{sec:results}
We first outline the experimental setup. Next, we evaluate both interval and affine arithmetic QIBP on several different datasets and QML models. Next, we explore the effect of different hyperparameters and  loss functions on the resulting certified accuracy in relation to the original test accuracy. 

\subsection{Experimental Setup}
We use Python v3.13.5, PennyLane v0.42.3~\cite{bergholm2022pennylane}, PyTorch v2.11.0~\cite{ansel2024pytorch}, and torchvision v0.26.0 to carry out training and experimental evaluation of the trained models. All experiments are performed on NVIDIA B200 and NVIDIA L4 GPUs.

\subsubsection{Datasets}
The VQCs were trained on MNIST~\cite{lecun1998mnist}, FashionMNIST (FMNIST)~\cite{xiao2017fashionmnist}, and Kuzushiji-MNIST (KMNIST)~\cite{clanuwat2018deep} datasets. These datasets each contain 60,000 training images and 10,000 testing images, where each image is grayscale of size $28\times28$. We rescale the images appropriately for the number of qubits in the model under amplitude embedding. For example, an 8 qubit model will resize the images to $16\times16$. Additionally, some models used less classes than the 10 that is provided by default with each dataset. For these classes, we use the labels in the range of the number of classes. For example, if the model classifies 4 classes, we select the data samples with class labels of 0 to 3.

\subsubsection{Training Hyperparameters}
To train the VQCs using QIBP with the loss functions as defined in Equations~\ref{eq:loss} and~\ref{eq:marginloss}, we utilize varying hyperparameter values. All optimizations in the training routines are carried out with the Adam optimizer~\cite{kingma2014adam} with a weight decay of 0.01 and a learning rate of 0.005 for 30 epochs. These 30 epochs are divided into different sections to adjust $\kappa$. We set $\kappa=1.0$ for the first 5 epochs to let the model learn clean classification first. In the next 15 epochs, $\kappa$ decreases at a constant rate until it reaches a target $\kappa$, such as $\kappa=0.50$. The remaining epochs are done at the resulting $\kappa$ value. We also use this same notion for scheduling $\epsilon$, where $\epsilon$ starts at $\epsilon=0.000$ and increases at a constant rate over 15 epochs to a target $\epsilon$ value.



\subsubsection{Metrics}
To evaluate the effectiveness of QIBP in training the models for robust decision boundaries, three different accuracy metrics are used. Test accuracy is the traditional accuracy of the model on the test dataset, representing the correctly classified data samples out of the total. Certified accuracy represents the worst case lower bound on accuracy for the model under the given perturbation $\epsilon$. To certify a sample, the lower bound logit of the correct class label must be larger than the upper bound logits of the incorrect class labels, like is done in the loss function (see Section~\ref{sec:loss}). PGD accuracy is the number of correct labels that were predicted from adversarial samples, crafted with a PGD attack of similar $\epsilon$ strength. This represents the upper bound for the trained model. Ideally, all three accuracies will be tightly bounded for robust decision boundaries.



\subsection{Accuracies from Training with Interval Arithmetic}
We first present the test accuracy and the certified accuracy on QIBP trained models using interval arithmetic. These models are able to train to sufficient test accuracy and in some model configurations, are bounded tightly between the certified accuracy and the test accuracy, as shown in Table~\ref{tab:interval} and Figure~\ref{fig:interval}. The tightness between the certified accuracy and the test accuracy typically occurs with the shallow and narrow models, such as MNIST 4 qubits, 2 classes, and 2 layers with margin loss, where the difference between the test and certified accuracy is $0.99\%$. This additionally includes the PGD accuracy, which is tightly bounded to both the certified and test accuracies, indicating that the interval QIBP trained model established robust decision boundaries, even in the worst case as represented with the propagated intervals throughout the model.

However, the certified accuracies begin decreasing rapidly compared to test accuracy as the models grow in complexity. In the worst case, we experience a difference of $39.71\%$ in MNIST 10 qubits, 2 classes, and 2 layers with margin loss. Despite this large difference between the test and certified accuracies, the PGD accuracy has a difference of $0.14\%$, demonstrating that the model can make correct predictions despite adversarial samples. The certified accuracy, representing the worst case lower bound, can result in high variations depending on the adversarial samples given under $\epsilon$ perturbation. This is likely due to interval arithmetic covering a wide range of values with a rectangle instead of adapting to a different shape for tighter bounds. With affine arithmetic, we can tighten the bounds and achieve tighter bounded certified accuracies, as discussed in Section~\ref{sec:aff-results}.

In terms of the loss functions, margin loss typically produced more accurate models in regards to each of the three metrics observed. Cross entropy loss saw better models in some of the increased qubit and class count models, however the typical increase margin loss obtains compared to these few cases provides greater benefits. Therefore, margin loss should be chosen as the loss function in interval arithmetic.


\begin{table}
\caption{Fixed $\epsilon=0.001$ and $\kappa=0.5$ test accuracy (Acc.) and certified accuracy (Cert.) results across datasets (D), cross entropy and margin loss functions, and different model configurations in terms of qubits (Q), classes (C), and layers (L) for interval arithmetic. Additionally, the accuracy on PGD attack is given for the empirical upper bound.}
\label{tab:interval}
\begin{tabular}{cccccccccc}
\toprule
 \multirow{2}{*}{\textbf{D}} & \multirow{2}{*}{\textbf{Q}} & \multirow{2}{*}{\textbf{C}} & \multirow{2}{*}{\textbf{L}} & \multicolumn{3}{c}{\textbf{Cross Entropy Loss}} & \multicolumn{3}{c}{\textbf{Margin Loss}} \\
 \cmidrule(lr){5-7} \cmidrule(lr){8-10}
 &  &  &  & \textbf{Acc.} & \textbf{Cert.} & \textbf{PGD} & \textbf{Acc.} & \textbf{Cert.} & \textbf{PGD} \\
\midrule
\parbox[t]{2mm}{\multirow{16}{*}{\rotatebox[origin=c]{90}{MNIST}}} & \multirow{4}{*}{4} & \multirow{2}{*}{2} & 2 & 92.48 & 90.50 & 92.10 & 96.36 & 95.37 & 96.12 \\
 &  &  & 8 & 94.00 & 92.01 & 93.81 & 97.07 & 94.37 & 96.78 \\
\cmidrule{3-10}
 &  & \multirow{2}{*}{4} & 2 & 45.30 & 38.06 & 44.07 & 66.63 & 51.89 & 64.57 \\
 &  &  & 8 & 75.58 & 65.19 & 74.62 & 82.73 & 72.17 & 81.72 \\
\cmidrule{2-10} \cmidrule{3-10}
 & \multirow{4}{*}{6} & \multirow{2}{*}{2} & 2 & 98.58 & 97.92 & 98.49 & 99.24 & 98.58 & 99.20 \\
 &  &  & 8 & 98.96 & 97.59 & 98.91 & 99.57 & 98.11 & 99.57 \\
\cmidrule{3-10}
 &  & \multirow{2}{*}{6} & 2 & 41.20 & 30.58 & 39.83 & 44.79 & 29.25 & 42.65 \\
 &  &  & 8 & 44.35 & 29.53 & 43.72 & 59.01 & 34.22 & 58.23 \\
\cmidrule{2-10} \cmidrule{3-10}
 & \multirow{4}{*}{8} & \multirow{2}{*}{2} & 2 & 95.37 & 87.19 & 95.04 & 98.11 & 95.22 & 97.97 \\
 &  &  & 8 & 96.74 & 79.72 & 96.50 & 99.48 & 69.46 & 99.43 \\
\cmidrule{3-10}
 &  & \multirow{2}{*}{8} & 2 & 27.53 & 13.68 & 26.79 & 27.23 & 12.71 & 26.51 \\
 &  &  & 8 & 26.82 & 9.54 & 25.89 & 24.59 & 8.03 & 23.31 \\
\cmidrule{2-10} \cmidrule{3-10}
 & \multirow{4}{*}{10} & \multirow{2}{*}{2} & 2 & 96.88 & 55.65 & 96.74 & 98.01 & 58.30 & 97.87 \\
 &  &  & 8 & 97.73 & 67.75 & 97.59 & 98.01 & 80.99 & 97.78 \\
\cmidrule{3-10}
 &  & \multirow{2}{*}{10} & 2 & 10.41 & 0.41 & 8.92 & 10.38 & 0.42 & 8.92 \\
 &  &  & 8 & 8.93 & 1.68 & 8.15 & 8.95 & 1.55 & 8.06 \\
\cmidrule{1-10} \cmidrule{2-10} \cmidrule{3-10}
\parbox[t]{2mm}{\multirow{16}{*}{\rotatebox[origin=c]{90}{FMNIST}}} & \multirow{4}{*}{4} & \multirow{2}{*}{2} & 2 & 91.95 & 89.20 & 91.85 & 92.45 & 89.70 & 92.30 \\
 &  &  & 8 & 94.70 & 93.25 & 94.20 & 94.45 & 92.85 & 94.30 \\
\cmidrule{3-10}
 &  & \multirow{2}{*}{4} & 2 & 55.00 & 53.40 & 54.75 & 56.08 & 51.15 & 55.60 \\
 &  &  & 8 & 82.83 & 69.33 & 82.28 & 82.97 & 74.53 & 82.47 \\
\cmidrule{2-10} \cmidrule{3-10}
 & \multirow{4}{*}{6} & \multirow{2}{*}{2} & 2 & 91.85 & 89.20 & 91.55 & 92.45 & 88.80 & 92.15 \\
 &  &  & 8 & 92.50 & 87.50 & 92.30 & 93.50 & 89.15 & 93.30 \\
\cmidrule{3-10}
 &  & \multirow{2}{*}{6} & 2 & 26.70 & 15.95 & 26.20 & 41.53 & 14.63 & 40.30 \\
 &  &  & 8 & 45.83 & 40.17 & 45.45 & 57.92 & 39.97 & 56.98 \\
\cmidrule{2-10} \cmidrule{3-10}
 & \multirow{4}{*}{8} & \multirow{2}{*}{2} & 2 & 51.85 & 49.55 & 51.60 & 91.75 & 74.90 & 91.55 \\
 &  &  & 8 & 93.85 & 85.85 & 93.80 & 94.35 & 87.05 & 94.20 \\
\cmidrule{3-10}
 &  & \multirow{2}{*}{8} & 2 & 24.41 & 5.31 & 24.23 & 21.82 & 11.56 & 19.20 \\
 &  &  & 8 & 27.48 & 19.06 & 27.09 & 28.84 & 15.14 & 27.90 \\
\cmidrule{2-10} \cmidrule{3-10}
 & \multirow{4}{*}{10} & \multirow{2}{*}{2} & 2 & 74.50 & 37.15 & 73.70 & 83.40 & 43.15 & 82.95 \\
 &  &  & 8 & 58.90 & 46.85 & 58.20 & 93.10 & 70.25 & 92.95 \\
\cmidrule{3-10}
 &  & \multirow{2}{*}{10} & 2 & 19.95 & 0.70 & 19.26 & 11.25 & 0.00 & 6.80 \\
 &  &  & 8 & 9.90 & 0.00 & 7.69 & 13.84 & 0.03 & 9.62 \\
\cmidrule{1-10} \cmidrule{2-10} \cmidrule{3-10}
\parbox[t]{2mm}{\multirow{16}{*}{\rotatebox[origin=c]{90}{KMNIST}}} & \multirow{4}{*}{4} & \multirow{2}{*}{2} & 2 & 92.80 & 91.45 & 92.65 & 92.95 & 91.65 & 92.80 \\
 &  &  & 8 & 94.40 & 91.90 & 94.30 & 95.80 & 92.90 & 95.70 \\
\cmidrule{3-10}
 &  & \multirow{2}{*}{4} & 2 & 56.95 & 50.45 & 56.02 & 57.05 & 49.12 & 55.83 \\
 &  &  & 8 & 64.75 & 45.20 & 63.75 & 69.12 & 47.67 & 68.38 \\
\cmidrule{2-10} \cmidrule{3-10}
 & \multirow{4}{*}{6} & \multirow{2}{*}{2} & 2 & 92.10 & 84.70 & 91.60 & 93.10 & 83.80 & 92.75 \\
 &  &  & 8 & 92.25 & 86.65 & 92.10 & 94.05 & 85.65 & 93.75 \\
\cmidrule{3-10}
 &  & \multirow{2}{*}{6} & 2 & 30.82 & 20.52 & 30.03 & 34.03 & 22.78 & 33.20 \\
 &  &  & 8 & 38.30 & 13.08 & 37.42 & 38.70 & 13.77 & 37.75 \\
\cmidrule{2-10} \cmidrule{3-10}
 & \multirow{4}{*}{8} & \multirow{2}{*}{2} & 2 & 87.95 & 66.50 & 87.70 & 90.25 & 68.75 & 89.95 \\
 &  &  & 8 & 91.65 & 80.25 & 91.35 & 91.00 & 35.30 & 90.45 \\
\cmidrule{3-10}
 &  & \multirow{2}{*}{8} & 2 & 15.57 & 5.99 & 14.82 & 14.36 & 3.65 & 13.44 \\
 &  &  & 8 & 22.06 & 13.18 & 21.43 & 18.86 & 6.21 & 17.95 \\
\cmidrule{2-10} \cmidrule{3-10}
 & \multirow{4}{*}{10} & \multirow{2}{*}{2} & 2 & 74.50 & 48.50 & 73.70 & 76.90 & 48.30 & 75.90 \\
 &  &  & 8 & 85.70 & 39.60 & 85.45 & 85.20 & 43.15 & 84.90 \\
\cmidrule{3-10}
 &  & \multirow{2}{*}{10} & 2 & 10.79 & 0.21 & 9.00 & 10.78 & 0.20 & 8.99 \\
 &  &  & 8 & 9.93 & 0.96 & 8.83 & 9.89 & 0.94 & 8.86 \\
\bottomrule
\end{tabular}
\end{table}

\subsection{Increasing Accuracies with Affine Arithmetic Training} \label{sec:aff-results}
\begin{figure}
    \begin{subfigure}{\linewidth}
        \centering
        \includegraphics[width=\linewidth]{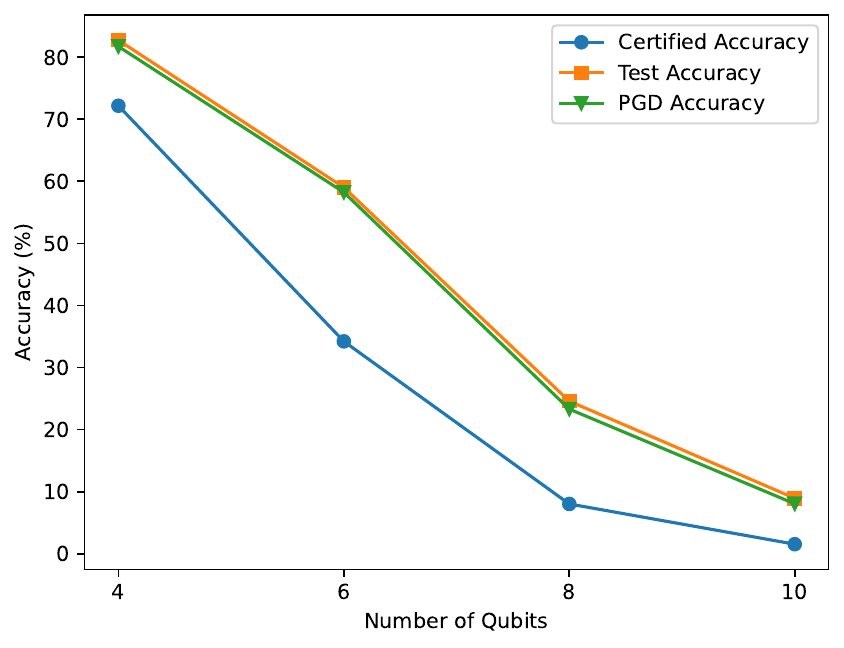}
        \vspace{-0.2in}
        \caption{Interval arithmetic.}
        \label{fig:interval}
    \end{subfigure}
    
    \vspace{0.1in}
    
    \begin{subfigure}{\linewidth}
        \centering
        \includegraphics[width=\linewidth]{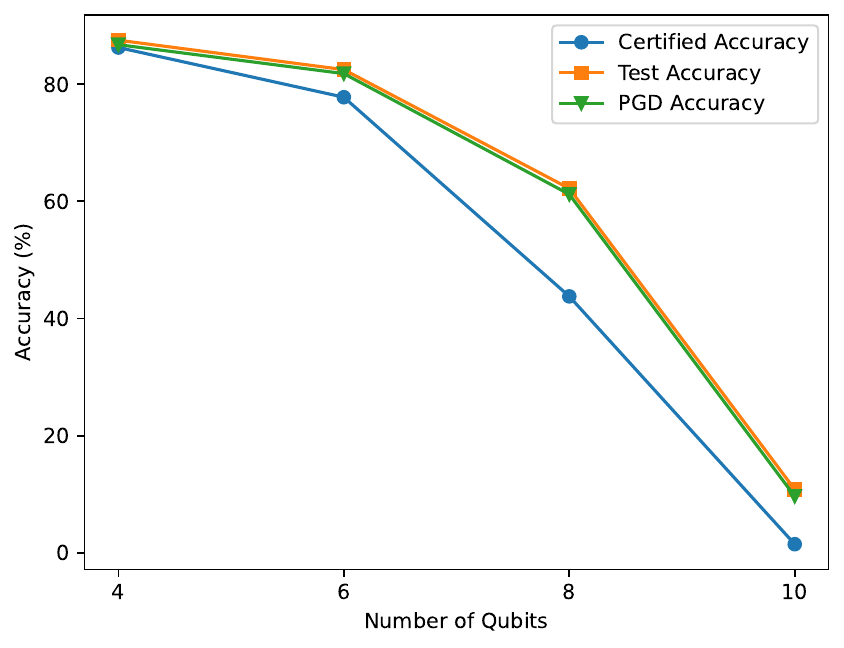}
        \vspace{-0.2in}
        \caption{Affine arithmetic.}
        \label{fig:affine}
    \end{subfigure}
    \caption{Accuracies resulting from (a) interval arithmetic and (b) affine arithmetic QIBP. All models shown are of varying qubits, maximum classes, and 8 layers under margin loss.}
    \label{fig:comparison}
    \vspace{-0.1in}
\end{figure}

We evaluate the resulting certified and test accuracies of different model configurations trained with affine arithmetic QIBP. These accuracies are shown in Table~\ref{tab:major} and Figure~\ref{fig:affine}. We analyze the results at a fixed $\epsilon=0.001$ and $\kappa=0.5$ to understand how model features such as the number of qubits, layers, and classes affects the resulting certified accuracy. Additionally, with the cross entropy loss function (Equation~\ref{eq:loss}) and the margin loss (Equation~\ref{eq:marginloss}), there can be differences in the resulting accuracies.

For the MNIST dataset, when the model is trained with the cross entropy loss function, both the test accuracy and the certified accuracy achieve high values under small qubit and class constraints. The MNIST VQC model over 4 qubits classifying 2 classes with 8 layers was able to reach $94.61\%$ test accuracy and $94.42\%$ certified accuracy. As the number of qubits grow, both the test accuracy and the certified accuracy decrease, and most notably, the difference between the test accuracy and the certified accuracy increasing. At 10 qubits classifying 2 classes in 8 layers, the test accuracy is $98.11\%$ and certified accuracy of $92.34\%$.

However, as the classes increase, both the test accuracy and the certified accuracy drastically drop, as the classification problem becomes more complex and the loss function has to optimize more bounds due to increasing number of logits, and thus, logits representing incorrect classes that must have an upper bound that is lower than the lower bound of the correct class logit. With 10 qubits, 10 classes, and 8 layers, the resulting accuracies are $20.82\%$ and $13.83\%$, respectively. While the accuracies do decrease, the difference between the test accuracy and the certified accuracy remain similar to that under 2 classes, where 2 classes had a difference of $5.77\%$ and 10 classes a difference of $6.99\%$. 

Generally, this is the trend where as the model grows more complex in terms of size with the number of qubits, classes, and layers, both the test accuracy and the certified accuracy decrease. For example, with 6 qubits, 6 classes, and 8 layers, the accuracy with the cross-entropy loss is $72.11\%$ for test accuracy and $67.63\%$ for certified accuracy. Under margin loss, these accuracies improve to $82.47\%$ and $77.76\%$, respectively, with the difference between the two accuracies also improving. This indicates that for this specific model configuration, the margin loss penalty was successful in having the model formulate better decision bounds for the class labels. 

While the margin loss does improve many model configurations compared to the cross entropy loss, there are instances where the resulting training produces lower test and certified accuracies, indicating a potential issue with learning effective bounds. This is the case with the 10 qubit models, where the test accuracy drops to $10.84\%$ and certified to $1.50\%$ for 10 classes and 8 layers. Therefore, while generally the margin loss does help improve both accuracies, there are some cases where it can perform worse, most likely due to the resulting bounds and logits acquired for the data samples during training. Similar trends appear with the resulting test and certified accuracies for FMNIST and KMNIST. These datasets are more complex compared to MNIST, meaning that their resulting accuracies will be lower compared to MNIST with the same model configurations. Despite this increased complexity from the data itself, the same accuracy trends are observed.

These accuracies are also compared with the accuracy obtained by running a comparable strength PGD attack on the model. As shown in Table~\ref{tab:major}, for smaller models, the difference between the certified and PGD accuracies is small, indicating a tight bound between the worst case accuracy and the upper bound. With the model increasing in size and classes, this difference increases, however it still maintains a difference of $18.41\%$ in the worst case for MNIST models, occurring with 8 qubits, 8 classes, and 8 layers.

\begin{table}[]
    \centering
    \caption{Fixed $\epsilon=0.001$ and $\kappa=0.5$ test accuracy (Acc.) and certified accuracy (Cert.) results across datasets (D), cross entropy and margin loss functions, and different model configurations in terms of qubits (Q), classes (C), and layers (L) for affine arithmetic. Additionally, the accuracy on PGD attack is given for the empirical upper bound.}
    \label{tab:major}
    \begin{tabular}{cccccccccc}
        \toprule
        \multirow{2}*{\textbf{D}} & \multirow{2}*{\textbf{Q}} & \multirow{2}*{\textbf{C}} & \multirow{2}*{\textbf{L}} & \multicolumn{3}{c}{\textbf{Cross Entropy Loss}} & \multicolumn{3}{c}{\textbf{Margin Loss}}\\ 
        \cmidrule(lr){5-7} \cmidrule(lr){8-10}
        & & & & \textbf{Acc.} & \textbf{Cert.} & \textbf{PGD} & \textbf{Acc.} & \textbf{Cert.} & \textbf{PGD}\\
        \midrule
        \parbox[t]{2mm}{\multirow{16}{*}{\rotatebox[origin=c]{90}{MNIST}}} & \multirow{4}*{4} & \multirow{2}*{2} & 2 & 92.34 & 91.54 & 91.63 & 96.50 & 96.17 & 96.22\\
        & & & 8 & 94.61 & 94.42 & 94.52 & 97.12 & 96.69 & 96.83\\
        \cmidrule{3-10}
        & & \multirow{2}*{4} & 2 & 47.27 & 44.05 & 46.09 & 70.72 & 65.67 & 68.44\\
        & & & 8 & 80.92 & 79.50 & 79.94 & 87.49 & 86.24 & 86.72\\
        \cmidrule{2-10}
        & \multirow{4}*{6} & \multirow{2}*{2} & 2 & 98.77 & 98.20 & 98.68 & 99.24 & 99.05 & 99.24\\
        & & & 8 & 99.57 & 99.53 & 99.57 & 99.67 & 99.53 & 99.62\\
        \cmidrule{3-10}
        & & \multirow{2}*{6} & 2 & 45.53 & 40.28 & 44.47 &51.07 & 40.47 & 48.86\\
        & & & 8 & 72.11 & 67.63 & 71.27 & 82.47 & 77.76 & 81.79\\
        \cmidrule{2-10}
        & \multirow{4}*{8} & \multirow{2}*{2} & 2 & 96.83 & 94.18 & 96.64 &98.77 & 97.68 & 98.68\\
        & & & 8 & 99.67 & 99.43 & 99.67 & 99.76 & 99.43 & 99.76\\
        \cmidrule{3-10}
        & & \multirow{2}*{8} & 2 & 27.48 & 17.15 & 26.89 & 31.36 & 17.33 & 30.29\\
        & & & 8 & 57.35 & 42.56 & 56.53 & 62.19 & 43.78 &
        61.19\\
        \cmidrule{2-10}
        & \multirow{4}*{10} & \multirow{2}*{2} & 2 & 98.20 & 88.13 & 98.11 & 99.29 & 93.29 & 99.24\\
        & & & 8 & 98.11 & 92.34 & 97.92 & 99.24 & 93.57 & 99.10\\
        \cmidrule{3-10}
        & & \multirow{2}*{10} & 2 & 20.96 & 6.25 & 20.48 & 10.57 & 0.41 & 8.95\\
        & & & 8 & 20.82 & 13.83 & 20.49 & 10.84 & 1.50 & 9.67\\
        \midrule
        \parbox[t]{2mm}{\multirow{16}{*}{\rotatebox[origin=c]{90}{FMNIST}}} & \multirow{4}*{4} & \multirow{2}*{2} & 2 & 92.15 & 91.60 & 91.90 & 92.60 & 92.15 & 92.35\\
        & & & 8 & 94.65 & 94.10 & 94.45 & 94.65 & 94.15 & 94.40\\
        \cmidrule{3-10}
        & & \multirow{2}*{4} & 2 & 55.68 & 54.68 & 55.45 & 63.43 & 58.23 & 62.30\\
        & & & 8 & 82.43 & 80.25 & 82.03 & 83.55 & 82.03 & 83.08\\
        \cmidrule{2-10}
        & \multirow{4}*{6} & \multirow{2}*{2} & 2 & 92.10 & 91.35 & 91.80 & 92.50 & 91.15 & 92.15\\
        & & & 8 & 93.45 & 92.40 & 93.20 & 93.60 & 92.80 & 93.30\\
        \cmidrule{3-10}
        & & \multirow{2}*{6} & 2 & 27.57 & 22.43 & 26.83 & 48.52 & 29.98 & 46.58\\
        & & & 8 & 72.62 & 58.90 & 71.40 & 71.20 & 61.13 & 70.38\\
        \cmidrule{2-10}
        & \multirow{4}*{8} & \multirow{2}*{2} & 2 & 71.90 & 54.40 & 70.80 & 93.35 & 89.55 & 93.30\\
        & & & 8 & 94.65 & 93.55 & 94.50 & 94.85 & 93.80 & 94.85\\
        \cmidrule{3-10}
        & & \multirow{2}*{8} & 2 & 27.13 & 0.86 & 26.26 & 30.16 & 12.14 & 28.01\\
        & & & 8 & 47.85 & 17.30 & 46.74 & 58.86 & 30.46 & 56.88\\
        \cmidrule{2-10}
        & \multirow{4}*{10} & \multirow{2}*{2} & 2 & 90.15 & 67.55 & 89.75 & 89.55 & 72.35 & 89.30\\
        & & & 8 & 94.75 & 91.25 & 94.50 & 94.70 & 91.05 & 94.65\\
        \cmidrule{3-10}
        & & \multirow{2}*{10} & 2 & 20.40 & 0.66 & 19.90 & 11.15 & 0.00 & 6.69\\
        & & & 8 & 11.47 & 0.64 & 11.06 & 23.99 & 6.65 & 21.06\\
        \midrule
        \parbox[t]{2mm}{\multirow{16}{*}{\rotatebox[origin=c]{90}{KMNIST}}} & \multirow{4}*{4} & \multirow{2}*{2} & 2 & 92.60 & 92.30 & 92.45 & 93.10 & 92.60 & 92.75\\
        & & & 8 & 93.85 & 93.65 & 93.65 & 96.15 & 95.60 & 95.85\\
        \cmidrule{3-10}
        & & \multirow{2}*{4} & 2 & 54.53 & 51.93 & 53.30 & 55.68 & 52.78 & 54.53\\
        & & & 8 & 71.18 & 69.25 & 70.35 & 71.35 & 69.40 & 70.55\\
        \cmidrule{2-10}
        & \multirow{4}*{6} & \multirow{2}*{2} & 2 & 93.30 & 91.70 & 93.00 & 93.45 & 91.80 & 93.00\\
        & & & 8 & 94.75 & 93.80 & 94.65 & 96.50 & 95.60 & 96.30\\
        \cmidrule{3-10}
        & & \multirow{2}*{6} & 2 & 33.30 & 27.57 & 32.52 & 37.40 & 29.92 & 36.15\\
        & & & 8 & 52.00 & 45.88 & 50.93 & 59.17 & 52.60 & 58.22\\
        \cmidrule{2-10}
        & \multirow{4}*{8} & \multirow{2}*{2} & 2 & 89.05 & 85.20 & 88.70 & 91.00 & 85.85 & 90.35\\
        & & & 8 & 92.35 & 89.65 & 92.00 & 95.15 & 90.30 & 95.00\\
        \cmidrule{3-10}
        & & \multirow{2}*{8} & 2 & 22.26 & 7.98 & 21.23 & 19.06 & 5.89 & 17.29\\
        & & & 8 & 33.55 & 14.08 & 32.40 & 37.81 & 14.58 & 32.40 \\
        \cmidrule{2-10}
        & \multirow{4}*{10} & \multirow{2}*{2} & 2 & 83.60 & 67.15 & 83.05 & 83.70 & 67.20 & 83.00\\
        & & & 8 & 87.95 & 74.45 & 87.70 & 89.60 & 75.25 & 89.25\\
        \cmidrule{3-10}
        & & \multirow{2}*{10} & 2 & 10.83 & 0.20 & 9.03 & 10.82 & 0.20 & 9.02\\
        & & & 8 & 11.05 & 1.19 & 10.71 & 10.43 & 0.96 & 9.19\\
        \bottomrule
    \end{tabular}
\end{table}


\subsection{Exploration of Hyperparameter $\epsilon$}
We evaluate the effect of changing $\epsilon$ on the resulting accuracies for the different model configurations on the MNIST dataset, finding tradeoffs between the resulting accuracies and the robustness the model can handle with the specific value of $\epsilon$. We start with $\epsilon=0.001$ as shown in prior results and increase $\epsilon$ to $0.010$, as shown  in Table~\ref{tab:esp_sweep}.

The number of qubits the model operates on determines how detailed the input features are. With more qubits, the larger the image that can be used as input can be. As $\epsilon$ increases for each qubit value, the test accuracy remains similar while the certified accuracy decreases. For example, with 4 qubits, 2 classes, and 8 layers, the test accuracy ranges from minimum $98.97\%$ at $\epsilon=0.005$ to maximum $97.16\%$ at $\epsilon=0.010$. However, the certified accuracy drops as $\epsilon$ increases, going from $96.69\%$ at $\epsilon=0.001$ to $93.38\%$ at $\epsilon=0.010$. This drop in certified accuracy is largely due to the bounds calculated for each logit becoming looser, resulting in more training required to learn effective decision boundaries for similar certification accuracy as the lower $\epsilon$ values.

This trend is observed as the qubits increase. In addition, as the qubits increase, the certified accuracies also increase under small $\epsilon$. The certified accuracies go from $96.93\%$ under 4 qubits to $99.15\%$ under 8 qubits, for the best performing $\epsilon$ value. However, as the $\epsilon$ increases from the small values, the certified accuracies at larger qubits tend to perform worse compared to smaller qubit models, in addition to a worsening difference between the test accuracy and the certified accuracy. For 8 qubits, the certified accuracy of $99.15\%$ at $\epsilon=0.001$ decreases to $70.64\%$ at $\epsilon=0.010$. The test accuracy at $\epsilon=0.010$ is $98.49\%$, resulting in an accuracy difference of $27.85\%$. This large decrease in certified accuracy is due to having more features to track and certify, resulting in larger potential for the perturbation to worsen the interval bounds propagated throughout the model and the final result from full propagation. Overall, larger qubit counts improve peak performance but appear more sensitive to larger $\epsilon$, indicating a trade-off between capacity and robustness stability across $\epsilon$ values.






As shown under fixed $\epsilon$, the number of classes can have a large impact on the resulting accuracies, primarily in the difference between the test accuracy and the certified accuracy. For 2 classes and $\epsilon$ values under 6 qubits and 4 layers, the test accuracy remains similar across increasing $\epsilon$, ranging from $98.87\%$ to $99.53\%$. Certified accuracy does decrease slightly, from $98.68\%$ at $\epsilon=0.001$ to $94.52\%$ at $\epsilon=0.010$, indicating that the model bounds decrease slightly in robustness as the input is more perturbed. 

When the classes increase, both the test accuracy and the certified accuracy significantly drop, showcasing that the model has difficulties in processing more classes compared to the binary classification place. The difference between the test accuracy and certified accuracy is thus important to understand the impact on the robustness of the model bounds. For 4 classes, the certified accuracy falls from $85.21\%$ to $4.62\%$, while the test accuracy slightly decreases from $88.77\%$ to $87.52\%$ as the classes increase from 2 to 6. There is a large discrepancy between the differences of the test accuracy and the certified accuracy as the classes increase, going from a difference of $3.26\%$ at $\epsilon=0.001$ to a difference of $82.90\%$ at $\epsilon=0.010$. The decision boundaries are less robust as the $\epsilon$ increases, indicating difficulties in establishing boundaries that are robust against the $\epsilon$ perturbation that satisfies the bound constraints (Equation~\ref{eq:logits}).






Another important aspect is the number of layers in the model. As the number of layers increases, the corresponding test accuracy also increases. This increase in the test accuracy continues or remains similar as $\epsilon$ increases. Even at the highest $\epsilon=0.010$, with 4 qubits the test accuracy increases from $94.85\%$ at 2 layers to $97.16\%$ at 8 layers. Certified accuracy tends to decrease as $\epsilon$ grows within a layer, however, as the layers increase, the certified accuracy decrease tends to become smaller. For example, with 2 layers, the certified accuracy drops from $96.17\%$ to $84.44\%$ as $\epsilon$ increases from $0.001$ to $0.010$. Under 8 layers, the certified accuracy decreases from $96.69\%$ at $\epsilon=0.001$ to $93.38\%$ at $\epsilon=0.010$, indicating that more layers allows for better feature learning to establish more robust decision boundaries from training. Increasing the number of layers thus improves test accuracy but can have limited impact on improving certified accuracy as $\epsilon$ increases.

\begin{table}[]
    \centering
    \caption{Results for different $\epsilon$ values for different MNIST model configurations with margin loss. $\kappa$ is set to $0.50$.}
    \label{tab:esp_sweep}
    \begin{tabular}{cccccc}
        \toprule
        \textbf{Qubits} & \textbf{Classes} & \textbf{Layers} & \textbf{$\epsilon$} & \textbf{Acc.} & \textbf{Cert. Acc.} \\
        \midrule
         \multirow{12}*{4} & \multirow{12}*{2} & \multirow{3}*{2} & 0.001 & 96.50 & 96.17\\
         & & & 0.005 &95.84 & 92.06 \\
         & & & 0.010 & 94.85 & 84.44\\
         \cmidrule{3-6}
         & & \multirow{3}*{4} & 0.001 & 97.21 & 96.93\\
         & & & 0.005 & 97.16 & 95.56\\
         & & & 0.010 & 96.97 & 93.66\\
         \cmidrule{3-6}
         & & \multirow{3}*{6} & 0.001 & 97.30 & 96.93\\
         & & & 0.005 & 97.40 & 95.79\\
         & & & 0.010 & 97.02 & 93.43 \\
         \cmidrule{3-6}
         & & \multirow{3}*{8} & 0.001 & 97.12 & 96.69\\
         & & & 0.005 & 96.97 & 95.32\\
         & & & 0.010 & 97.16 & 93.38\\
         \midrule
         \multirow{9}*{6} & \multirow{3}*{2} & \multirow{3}*{4} & 0.001 & 98.87 & 98.68\\
         & & & 0.005 & 99.34 & 97.78 \\
         & & & 0.010 & 99.53 & 94.52\\
         \cmidrule{2-6}
         & \multirow{3}*{4} & \multirow{3}*{4} & 0.001 & 88.77 & 85.21\\
         & & & 0.005 & 88.04 & 56.72\\
         & & & 0.010 & 87.52 & 4.62 \\
         \cmidrule{2-6}
         & \multirow{3}*{6} & \multirow{3}*{4} & 0.001 & 77.72 & 70.68\\
         & & & 0.005 & 74.66 & 31.02\\
         & & & 0.010 & 71.51 & 1.76\\
         \midrule
         \multirow{3}*{8} & \multirow{3}*{2} & \multirow{3}*{4} & 0.001 & 99.62 & 99.15\\
         & & & 0.005 & 99.29 & 93.14\\
         & & & 0.010 & 98.49 & 70.64\\
         \bottomrule
    \end{tabular}
\end{table}

\subsection{Exploration of Hyperparameter $\kappa$}
Another hyperparameter that can be tuned is $\kappa$. Like in the previous ablation studies, we investigate the impact that changing $\kappa$ has on the test and certified accuracies under different model configurations with the MNIST dataset.

From Table~\ref{tab:k_sweep}, decreasing $\kappa$ from $0.9$ to $0.1$ can have varying effects on the resulting accuracies of the models. For 4 qubits, both accuracies increased, such as the certified accuracy increasing from $52.23\%$ at $\kappa=0.9$ to $67.55\%$ at $\kappa=0.1$ under 2 layers. The difference between the accuracies is similar as $\kappa$ decreases, indicating that improving the test accuracy also improves the certified accuracy at equivalent rates. However, as the qubits increase, this observed trend becomes less prevalent, instead dropping the test accuracy as $\kappa$ decreases. With 8 qubits, the test accuracy decreases from $68.52\%$ at $\kappa=0.9$ to $55.04\%$ at $\kappa=0.1$. The accuracies have a difference of $31.28\%$ at $\kappa=0.9$, which decreases to $10.44\%$ as $\kappa$ decreases to $0.1$. The resulting loss function is more sensitive to how much each loss term is contributing as $\kappa$ changes, resulting in a tighter certified accuracy when the worst case logit is the focus loss term with a small $\kappa$ compared to high test accuracy when the traditional classification loss term is the focus with a high $\kappa$.




Decreasing $\kappa$ improves both the test accuracy and the certified accuracy for each class value. While at 2 classes, the improvement is minimal due to the high accuracy at $\kappa=0.9$ already, it is beneficial as the classes increase. For 6 qubits and 4 classes, the test accuracies improve from $84.22\%$ to $88.43\%$ and the certified accuracies from $79.67\%$ to $84.97\%$ at $\kappa=0.9$ and $\kappa=0.1$, respectively. 6 classes features a similar trend, but with overall lower test and certified accuracies as the loss function must optimize more logit bounds with differing importance on its corresponding loss term in the loss function by the value of $\kappa$. The difficulty of the classification thus influences the impact that $\kappa$ has on the resulting accuracies of the model from training.





In regards to the number of layers in the model, as $\kappa$ decreases, both test and certified accuracies improve across all layer values. For shallower models such as with 2 layers, test accuracy improves from $56.22\%$ at $\kappa=0.9$ to $73.01\%$ at $\kappa=0.1$ and certified accuracy from $52.23\%$ to $67.55\%$. This indicates that smaller $\kappa$ values help stabilize training and improve robustness in shallow models, at the cost of  minimally increasing the difference between certified accuracy and test accuracy, from $3.99\%$ to $5.46\%$.

For deeper models, a similar trend is observed, although at more minimal improvements as $\kappa$ decreases. Reducing $\kappa$ from $0.9$ to $0.5$ yields large gains in both accuracies, however, any decrease in $\kappa$ below $0.5$ results in minimal improvements. With the 6 layer model, this is evidenced by the test accuracy increasing from $80.44\%$ at $\kappa=0.9$ to $89.01\%$ at $\kappa=0.5$, while barely improving to $89.51\%$ at $\kappa=0.1$. The certified accuracies exhibit a similar improvement pattern as $\kappa$ decreases. Therefore, the depth of the model in terms of the number of layers can improve the accuracies up to a point as $\kappa$ decreases. Importantly, $k=0.5$ consistently achieves strong performance across all layer settings and lies near the transition point where most of the performance gains have already been obtained, making it a reasonable choice.




\begin{table}[]
    \centering
    \caption{Results for different $\kappa$ values for different MNIST model configurations with margin loss. $\epsilon$ is set to 0.001.}
    \label{tab:k_sweep}
    \begin{tabular}{cccccc}
        \toprule
        \textbf{Qubits} & \textbf{Classes} & \textbf{Layers} & \textbf{$\kappa$} & \textbf{Acc.} & \textbf{Cert. Acc.} \\
        \midrule
         \multirow{12}*{4} & \multirow{12}*{4} & \multirow{3}*{2} & 0.9 & 56.22 & 52.23\\
         & & & 0.5 & 70.72 & 65.67 \\
         & & & 0.1 & 73.01 & 67.55\\
         \cmidrule{3-6}
         & & \multirow{3}*{4} & 0.9 & 75.90 & 73.59\\
         & & & 0.5 & 75.90 & 73.59\\
         & & & 0.1 & 82.80 & 80.32\\
         \cmidrule{3-6}
         & & \multirow{3}*{6} & 0.9 & 80.44 & 78.52\\
         & & & 0.5 & 89.01 & 87.25\\
         & & & 0.1 & 89.51 & 87.49 \\
         \cmidrule{3-6}
         & & \multirow{3}*{8} & 0.9 & 83.06 & 81.38\\
         & & & 0.5 & 87.49 & 86.24\\
         & & & 0.1 & 89.32 & 87.95\\
         \midrule
         \multirow{9}*{6} & \multirow{3}*{2} & \multirow{3}*{2} & 0.9 & 98.87 & 98.44\\
         & & & 0.5 & 99.24 & 99.05 \\
         & & & 0.1 & 99.53 & 99.34\\
         \cmidrule{2-6}
         & \multirow{3}*{4} & \multirow{3}*{2} & 0.9 & 84.22 & 79.67\\
         & & & 0.5 & 87.52 & 84.22\\
         & & & 0.1 & 88.43 & 84.97 \\
         \cmidrule{2-6}
         & \multirow{3}*{6} & \multirow{3}*{2} & 0.9 & 46.01 & 39.65\\
         & & & 0.5 & 51.07 & 40.47\\
         & & & 0.1 & 51.43 & 41.07\\
         \midrule
         \multirow{3}*{8} & \multirow{3}*{4} & \multirow{3}*{2} & 0.9 & 68.25 & 36.97\\
         & & & 0.5 & 57.42 & 41.62\\
         & & & 0.1 & 55.04 & 44.60\\
         \bottomrule
    \end{tabular}
\end{table}

\section{Conclusion} \label{sec:conc}
In this paper, we present a certified training technique for quantum machine learning (QML). Specifically, we propose quantum interval bound prorogation (QIBP) using both interval arithmetic and affine arithmetic to train QML models for adversarial robustness. QIBP is able to certify tightly to the model's test accuracy, guaranteeing perturbations up to $\epsilon$ will produce the correct class label, even in the worst case. Moreover, we show that the certified accuracy representing the worst case lower bound on accuracy is tight with the upper bound established by a comparable PGD attack.

\balance
\bibliographystyle{IEEEtran}
\bibliography{IEEEabrv,refs.bib}

\begin{thebibliography}{10}
\providecommand{\url}[1]{#1}
\csname url@samestyle\endcsname
\providecommand{\newblock}{\relax}
\providecommand{\bibinfo}[2]{#2}
\providecommand{\BIBentrySTDinterwordspacing}{\spaceskip=0pt\relax}
\providecommand{\BIBentryALTinterwordstretchfactor}{4}
\providecommand{\BIBentryALTinterwordspacing}{\spaceskip=\fontdimen2\font plus
\BIBentryALTinterwordstretchfactor\fontdimen3\font minus \fontdimen4\font\relax}
\providecommand{\BIBforeignlanguage}[2]{{%
\expandafter\ifx\csname l@#1\endcsname\relax
\typeout{** WARNING: IEEEtran.bst: No hyphenation pattern has been}%
\typeout{** loaded for the language `#1'. Using the pattern for}%
\typeout{** the default language instead.}%
\else
\language=\csname l@#1\endcsname
\fi
#2}}
\providecommand{\BIBdecl}{\relax}
\BIBdecl

\bibitem{schuld2015introduction}
M.~Schuld, I.~Sinayskiy, and F.~Petruccione, ``An introduction to quantum machine learning,'' \emph{Contemporary Physics}, vol.~56, no.~2, pp. 172--185, Apr. 2015.

\bibitem{lu2020quantum}
S.~Lu, L.-M. Duan, and D.-L. Deng, ``Quantum adversarial machine learning,'' \emph{Physical Review Research}, vol.~2, no.~3, p. 033212, Aug. 2020.

\bibitem{lecun1998mnist}
Y.~LeCun, ``The {{MNIST}} database of handwritten digits,'' 1998.

\bibitem{gowal2019scalable}
S.~Gowal, K.~D. Dvijotham, R.~Stanforth, R.~Bunel, C.~Qin, J.~Uesato, R.~Arandjelovic, T.~Mann, and P.~Kohli, ``Scalable {{Verified Training}} for {{Provably Robust Image Classification}},'' in \emph{Proceedings of the {{IEEE}}/{{CVF International Conference}} on {{Computer Vision}}}, 2019, pp. 4842--4851.

\bibitem{mao2024understanding}
Y.~Mao, M.~N. M{\"u}ller, M.~Fischer, and M.~Vechev, ``Understanding {{Certified Training}} with {{Interval Bound Propagation}},'' \emph{International Conference on Representation Learning}, vol. 2024, pp. 13\,470--13\,492, May 2024.

\bibitem{goodfellow2015explaining}
I.~J. Goodfellow, J.~Shlens, and C.~Szegedy, ``Explaining and {{Harnessing Adversarial Examples}},'' \emph{arXiv:1412.6572}, Mar. 2015.

\bibitem{madry2019deep}
A.~Madry, A.~Makelov, L.~Schmidt, D.~Tsipras, and A.~Vladu, ``Towards {{Deep Learning Models Resistant}} to {{Adversarial Attacks}},'' \emph{arXiv:1706.06083}, Sep. 2019.

\bibitem{pauli2022training}
P.~Pauli, A.~Koch, J.~Berberich, P.~Kohler, and F.~Allg{\"o}wer, ``Training {{Robust Neural Networks Using Lipschitz Bounds}},'' \emph{IEEE Control Systems Letters}, vol.~6, pp. 121--126, 2022.

\bibitem{khatun2025classical}
A.~Khatun and M.~Usman, ``Classical autoencoder distillation of quantum adversarial manipulations,'' \emph{Physical Review Research}, vol.~7, no.~4, p. L042054, Dec. 2025.

\bibitem{wendlinger2024comparative}
M.~Wendlinger, K.~Tscharke, and P.~Debus, ``A {{Comparative Analysis}} of {{Adversarial Robustness}} for {{Quantum}} and {{Classical Machine Learning Models}},'' in \emph{2024 {{IEEE International Conference}} on {{Quantum Computing}} and {{Engineering}} ({{QCE}})}, vol.~01, Sep. 2024, pp. 1447--1457.

\bibitem{berberich2024training}
J.~Berberich, D.~Fink, D.~Pranji{\'c}, C.~Tutschku, and C.~Holm, ``Training robust and generalizable quantum models,'' \emph{Physical Review Research}, vol.~6, no.~4, p. 043326, Dec. 2024.

\bibitem{perez-salinas2020data}
A.~{P{\'e}rez-Salinas}, A.~{Cervera-Lierta}, E.~{Gil-Fuster}, and J.~I. Latorre, ``Data re-uploading for a universal quantum classifier,'' \emph{Quantum}, vol.~4, p. 226, Feb. 2020.

\bibitem{lin2024veriqr}
Y.~Lin, J.~Guan, W.~Fang, M.~Ying, and Z.~Su, ``{{VeriQR}}: {{A Robustness Verification Tool}} for {{Quantum Machine Learning Models}},'' \emph{arXiv:2407.13533}, Jul. 2024.

\bibitem{assolini2025formal}
N.~Assolini, L.~Marzari, I.~Mastroeni, and A.~di~Pierro, ``Formal {{Verification}} of {{Variational Quantum Circuits}},'' \emph{arXiv:2507.10635}, Jul. 2025.

\bibitem{schuld2018supervised}
M.~Schuld and F.~Petruccione, \emph{Supervised {{Learning}} with {{Quantum Computers}}}, ser. Quantum {{Science}} and {{Technology}}.\hskip 1em plus 0.5em minus 0.4em\relax Springer International Publishing, 2018.

\bibitem{liu2016large}
W.~Liu, Y.~Wen, Z.~Yu, and M.~Yang, ``Large-margin softmax loss for convolutional neural networks,'' \emph{arXiv:1612.02295}, 2016.

\bibitem{zhang2021boosting}
B.~Zhang, D.~Jiang, D.~He, and L.~Wang, ``Boosting the certified robustness of l-infinity distance nets,'' \emph{arXiv:2110.06850}, 2021.

\bibitem{comba1993affine}
J.~L.~D. Comba and J.~Stolfi, ``Affine arithmetic and its applications to computer graphics,'' in \emph{Proceedings of {{VI SIBGRAPI}} (Brazilian, Symposium on Computer Graphics and Image Processing)}, 1993, pp. 9--18.

\bibitem{defigueiredo2004affine}
L.~H. {de Figueiredo} and J.~Stolfi, ``Affine {{Arithmetic}}: {{Concepts}} and {{Applications}},'' \emph{Numerical Algorithms}, vol.~37, no.~1, pp. 147--158, Dec. 2004.

\bibitem{bergholm2022pennylane}
V.~Bergholm \emph{et~al.}, ``{{PennyLane}}: {{Automatic}} differentiation of hybrid quantum-classical computations,'' \emph{arXiv:1811.04968}, Jul. 2022.

\bibitem{ansel2024pytorch}
J.~Ansel \emph{et~al.}, ``{{PyTorch}} 2: {{Faster Machine Learning Through Dynamic Python Bytecode Transformation}} and {{Graph Compilation}},'' in \emph{Proceedings of the 29th {{ACM International Conference}} on {{Architectural Support}} for {{Programming Languages}} and {{Operating Systems}}, {{Volume}} 2}, ser. {{ASPLOS}} '24, vol.~2.\hskip 1em plus 0.5em minus 0.4em\relax New York, NY, USA: Association for Computing Machinery, Apr. 2024, pp. 929--947.

\bibitem{xiao2017fashionmnist}
H.~Xiao, K.~Rasul, and R.~Vollgraf, ``Fashion-{{MNIST}}: A {{Novel Image Dataset}} for {{Benchmarking Machine Learning Algorithms}},'' \emph{arXiv:1708.07747}, Sep. 2017.

\bibitem{clanuwat2018deep}
T.~Clanuwat, M.~{Bober-Irizar}, A.~Kitamoto, A.~Lamb, K.~Yamamoto, and D.~Ha, ``Deep {{Learning}} for {{Classical Japanese Literature}},'' \emph{arXiv:1812.01718}, Nov. 2018.

\bibitem{kingma2014adam}
D.~P. Kingma and J.~Ba, ``Adam: {{A Method}} for {{Stochastic Optimization}},'' \emph{arXiv:1412.6980}, Dec. 2014.

\end{thebibliography}

\end{document}